# Directional Anisotropy of the Vibrational Modes in 2D Layered Perovskites

Balaji Dhanabalan[1,2] ‡, Yu-Chen Leng[3,4] ‡, Giulia Biffi[1,2], Miao-Ling Lin[3,4], Ping-Heng Tan[3,4], Ivan Infante[1], Liberato Manna[1], Milena P. Arciniegas[1], and Roman Krahne[1]*

[1]Istituto Italiano di Tecnologia (IIT), Via Morego 30, 16163 Genoa, Italy

[2]Dipartimento di Chimica e Chimica Industriale, Università degli Studi di Genova, Via Dodecaneso, 31, 16146 Genova, Italy

[3]State Key Laboratory of Superlattices and Microstructures, Institute of Semiconductors, Chinese Academy of Sciences, 100083 Beijing, China

[4]Center of Materials Science and Optoelectronics Engineering, University of Chinese Academy of Sciences, 100190 Beijing, China

**Abstract.** The vibrational modes in organic/inorganic layered perovskites are of fundamental importance for their optoelectronic properties. The hierarchical architecture of the Ruddlesden-Popper phase of these materials allows for distinct directionality of the vibrational modes with respect to the main axes of the pseudocubic lattice in the octahedral plane. Here, we study the directionality of the fundamental phonon modes in single exfoliated Ruddlesden-Popper perovskite flakes with polarized Raman spectroscopy at ultralow-frequencies. A wealth of Raman bands is distinguished in the range from 15-150 cm$^{-1}$ (2-15 meV), whose features depend on the organic cation species, on temperature, and on the direction of the linear polarization of the incident light. By controlling the angle of the linear polarization of the excitation laser with respect to the in-plane axes of the octahedral layer, we gain detailed information on the symmetry of the vibrational modes. The choice of two different organic moieties, phenethylammonium (PEA) and butylammonium (BA) allows to discern the influence of the linker molecules, evidencing strong anisotropy of the vibrations for the (PEA)$_2$PbBr$_4$ samples. Temperature dependent Raman measurements reveal that the broad phonon bands observed at room temperature consist of a series of sharp modes, and that such mode splitting strongly differs for the different organic moieties and vibrational bands. Softer molecules such as BA result in lower vibrational frequencies and splitting into fewer modes, while more rigid molecules such as PEA lead to higher frequency oscillations and larger number of Raman peaks at low temperature. Interestingly, in distinct bands the number of peaks in the Raman bands is doubled for the rigid PEA compared to the soft BA linkers. Our work shows that the coupling to specific vibrational modes can be controlled by the incident light polarization and choice of the organic moiety, which could be exploited for tailoring exciton-phonon interaction, and for optical switching of the optoelectronic properties of such 2D layered materials.

Two-dimensional (2D) Ruddlesden-Popper lead halide perovskites have emerged as a highly versatile material for optoelectronic applications[1,2] due to their bright emission,[3,4] strong excitonic and dielectric confinement,[5-7] and photostability in solar energy harvesting under ambient conditions.[8-11] This class of 2D layered perovskites can be fabricated using simple and low-cost strategies in the form of atomically thin layers,[12] single crystals,[13] and ensembles.[14,15] They consist of layers of corner-sharing $[PbX_6]^{4-}$ octahedra that are separated by organic ammonium-based cations (*e.g.* phenethylammonium (PEA) and butylammonium (BA)), which are too large to fit into a three-dimensional (3D) octahedral structure,[5,16,17] and form a natural superlattice of octahedral planes linked by interdigitated bilayers of the organic moieties. This hierarchical architecture has a variety of highly appealing properties that are different from their 3D framework: the inorganic octahedral layers form a quantum well potential for the electronic carriers, in which the confinement can be tuned by the number $n$ of adjacent octahedral planes,[15,16,18,19] an aspect that is also extremely interesting for fundamental studies;[13,20,21] the electron-phonon coupling (and distance) between the inorganic layers can be modified by the choice of the organic moiety;[14,22-24] and the electronic level structure as well as the band gap can be tailored by choice of the halide anion and *via* lattice deformations.[5,25-27] Furthermore, the presence of long hydrophobic organic moieties intercalated between the octahedral layers protects the layered perovskites from moisture permeation, conferring them structural and functional stability.[6,28,29]

The intrinsic structural orientation of the 2D Ruddlesden-Popper perovskites results in distinct optoelectronic and thermal properties due to the extremely different environments that the electrical, optical, and vibrational excitations encounter in- and out of plane of the layers.[20, 30] For example, the $[PbX_6]^{4-}$ layers behave as semiconductors, while the organic spacer is insulating.[31] Such structural anisotropy has also mechanical advantages: the photoluminescence of macroscopic stacks of 2D layered perovskites can be tuned by moderate external pressures due to the related anisotropy of the transition dipole moments,[32] and the layered structure allows for mechanical exfoliation of single flakes with a defined number of octahedral layers, and their transfer on suitable substrates.[14] The organic/inorganic bilayer structure behaves mechanically as an organic composite material reinforced by the octahedral layers, with out-of-plane weak bonds (van-der-Waals) and in-plane strong bonds (covalent or ionic), generating rigid cages when increasing the number of inorganic layers.[1, 33-35] Concerning the exciton binding energy, this hybrid structure has the peculiar consequence that the dielectric environment within the octahedral planes is similar to that of an inorganic crystal, leading to strongly bound 2D Wannier-Mott excitons,[36, 37] while the exciton confinement out-of-plane is defined by the organic moieties.[6] Such entangled exciton dynamics critically depend on coupling to phonons and distortions of the relatively soft lattice.[4, 38] Moreover, the exciton-phonon coupling strongly affects the scattering and decay channels of the excitons, and thus determines the absorbance and emission properties of the material.[13, 36, 39] Together with the thermal relaxation, these are essential for improving their current performance in optoelectronic devices, such as solar cells and light emitting diodes. Therefore, a detailed understanding of the properties of the vibrational resonances in layered perovskites is crucial to improve our knowledge on their orientation-dependent optoelectronic properties. Recent works already shed some light on the phonons in 2D layered Ruddlesden-Popper phase perovskites,[13, 39, 40] but a detailed experimental observation of the vibrational modes, in particular those in the ultralow-frequency regime, is still lacking.

In this work, we present a comprehensive study of the fundamental (ultralow-frequency) vibrational modes of single 2D $(PEA)_2PbBr_4$ and $(BA)_2PbBr_4$ layered perovskite flakes by polarized Raman spectroscopy. The flakes consist of multiple layers of single octahedral planes intercalated between the organic moieties ($n = 1$) and have lateral dimensions that typically exceed tens of micrometers. PEA and BA molecules were chosen as two types of organic moieties that differ in length, structure and symmetry to elucidate the influence of the organic spacer molecule (Figure 1). By correlating the main axes directions of the octahedral lattice with the linear polarization of the incident and detected light, we draw detailed conclusions on the directionality of different vibrational modes. In particular, the bands observed around 50 cm$^{-1}$ and 70 cm$^{-1}$, a frequency range that can be associated to Pb-Br bending modes, display very different symmetry, which is more pronounced for the (PEA)$_2$PbBr$_4$ flakes. Furthermore, these bands split into a large number of sharp peaks at low temperatures, and interestingly the number of peaks is doubled for the (PEA)$_2$PbBr$_4$ flakes with respect to the (BA)$_2$PbBr$_4$ ones, pointing to anisotropy effects or interlayer coupling. This different behavior highlights the importance of the organic moieties on the vibrational modes, and therefore on their impact on the electron-phonon interaction. Our work demonstrates that the vibrational modes of the 2D layered perovskites can be designed by the choice of organic linker molecule, and that different sets of modes can be preferentially excited by a proper choice of the angle of the linear polarization of the incident light with respect to the orientation of the octahedral lattice.

Results and Discussion

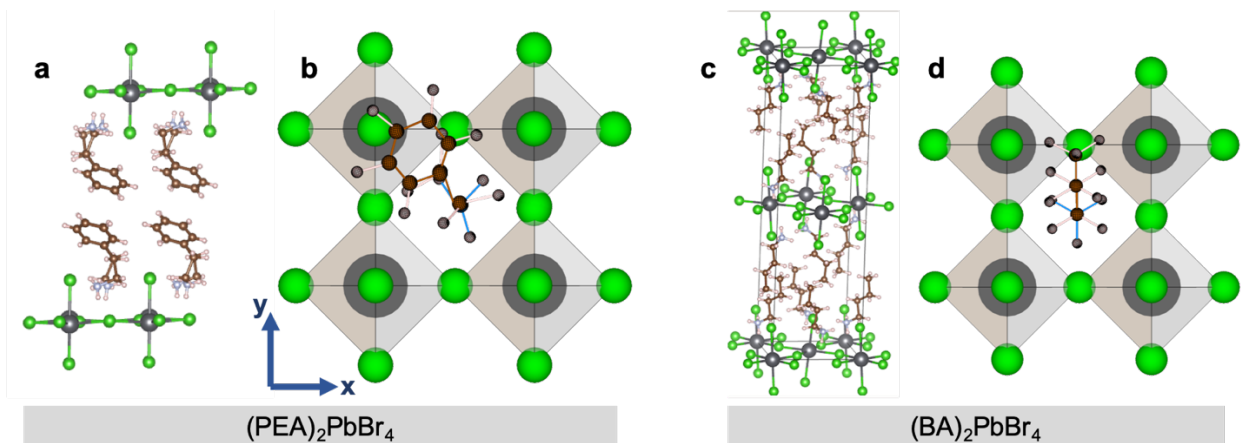

Figure 1. Structure of the organic and inorganic layers. (a,c) Unit cell of the lattice of the (PEA)$_2$PbBr$_4$ (a) and (BA)$_2$PbBr$_4$ (c) layered perovskites generated from the structure (CIF) files of ref. [13]. (b,d) Top view of the octahedral layer where the anchoring of the organic moiety is depicted. For the (PEA)$_2$PbBr$_4$ system the phenethyl ring is oriented in diagonal direction such that Pb – Br – π stacking occurs. The arrows in (b) indicate the *x.y* directions of the pseudocubic octahedral lattice. For simplicity, the octahedra are sketched without distortions.

The layered structure of the perovskite flakes with one octahedral layer (*n*=1) separated by the organic cation molecules is schematically shown in Figures 1 and 2a, and depicts the anchoring and coupling of the organic molecules. The conformation of the unit cells was plotted using the structure parameters that we obtained from our XRD analysis, which agree with those in ref.[13] In the top view in Figure 1b we oriented the bent PEA molecule in diagonal direction such that Pb – Br - π stacking is formed, which is motivated by our polarized Raman spectroscopy results, as will be discussed in detail later. While the PEA molecules contain phenethyl complexes that are expected to be relatively rigid, BA spacer molecules consist of a linear alkyl chain that is soft and more flexible. Therefore, torsional and bending oscillations of the molecules, [41, 42] induced by the vibrations of the inorganic lattice, are more pronounced in the PEA flakes, as confirmed by density functional theory (DFT) calculations. From the as-synthesized 2D layered perovskite materials single flakes (Figure 2b) can be exfoliated by using the scotch tape exfoliation technique (see Methods). These exfoliated flakes frequently manifest straight edges and rectangular corners as can be seen from the optical microscopy images in Figure S2. The X-ray diffraction patterns and optical characterization of the flakes are given in Figure S1 of the Supporting Information (SI), and show that the (001) planes are parallel to the substrate, with a spacing of 1.4 nm and 1.7 nm of the octahedral layers for $(BA)_2PbBr_4$ and $(PEA)_2PbBr_4$, respectively. These values correspond to *n = 1*, as we reported in our previous work.[14] Individual flakes manifest a single emission peak around 410 nm due to quantum confinement,[43] and evaluation of the height of the flake by atomic force microscopy allows to assign the number of organic/inorganic bilayers in the stack (see Figure S3).[14] Photoluminescence (PL) spectra of single $(PEA)_2PbBr_4$ flakes with different thicknesses are reported in Figure S4, and a slight redshift of the emission peak with increasing thickness can be observed, due most likely to self-absorption.

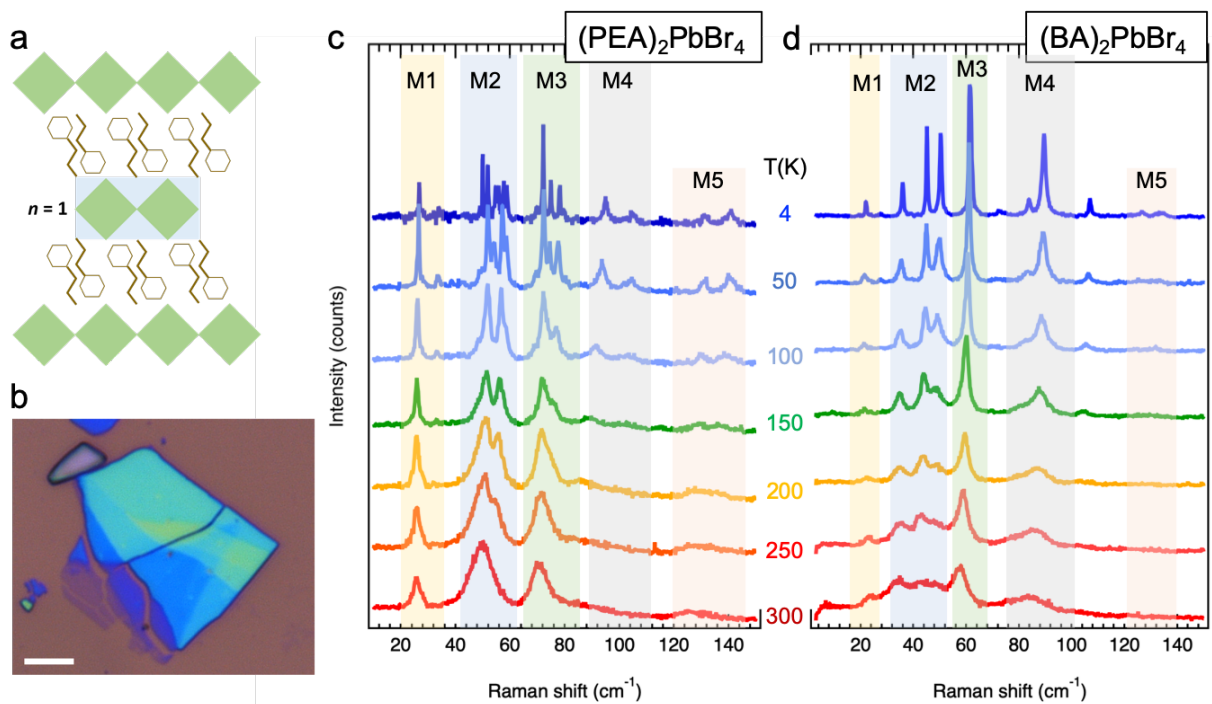

Figure 2. Raman spectra of exfoliated (PEA)$_2$PbBr$_4$ and (BA)$_2$PbBr$_4$ flakes. (a) Scheme of the 2D layered perovskite structure with PEA molecules as spacer and *n*=1. (b) Optical microscope image of exfoliated flakes with different thickness, where the different colors result from optical interferences. Scale bar is 5 μm. (c,d) Raman spectra of individual (PEA)$_2$PbBr$_4$ (c) and (BA)$_2$PbBr$_4$ (d) flakes recorded at different temperatures in the range from 4 to 300 K under laser excitation at 633 nm. The temperature is given for each spectrum, and the vibrational bands M1-M5 are indicated by the transparent colored rectangles.

Raman spectroscopy was performed under backscattering conditions at room temperature and in a helium-cooled cryostat (Montana Instruments), using a Jobin-Yvon HR800 micro-Raman system equipped with a liquid-nitrogen-cooled charge coupled device detector (CCD) and 50× objective lenses (N.A. = 0.45). An integrated CCD camera and a motorized sample stage allowed to select the regions of interest with micrometer precision. Raman spectra from single (PEA)$_2$PbBr$_4$ and (BA)$_2$PbBr$_4$ flakes recorded at room temperature show different vibrational bands in the ultralow-frequency regime (Figure 2c,d), and their frequencies do not depend on the number of layers of the measured flake, on the wavelength of the excitation laser (Figure S5), and are similar under ambient conditions or vacuum (Figure S6). Three dominant Raman bands (labeled as M1-M3) can be observed in the spectrum of the (PEA)$_2$PbBr$_4$ flakes at 300K (Figure 2c**)**, together with a band appearing as a broad shoulder at their high frequency side (M4), and another weak band (M5) around 125 cm$^{-1}$. These Raman bands are a convolution of several vibrational modes that gradually can be discerned with decreasing temperature. Interestingly, the different bands differ in terms of mode splitting, which should be closely related to the nature of the underlying vibrational modes. M1 remains a single peak that gradually becomes sharper with decreasing temperature, while M2 first splits into two peaks (in the range from 250-100 K), and then these peaks split again into doublets. A similar behavior is found for the M3 band, which indicates that these modes stem from similar lattice oscillations, which are most likely related to Pb-Br bond bending as these are dominant in this frequency range (see movies PEA_59cm-1.gif and PEA_76cm-1.gif in the Supplementary Material). The M4 band that appears as a broad shoulder on the high frequency side of the M3 band at room temperature develops into two well resolved peaks (at 95 cm$^{-1}$ and 105 cm$^{-1}$) at T=4 K. Finally, the M5 band around 125 cm$^{-1}$ splits up into two peaks at T=4 K that are located at 130 cm$^{-1}$ and 140 cm$^{-1}$. DFT modeling shows that oscillations in this frequency range mostly originate from Pb-Br bond stretching (see movie PEA_135cm-1.gif in the Supplementary Material). Although the main contributions of the vibrations of 2D-layered perovskites that we observe can be associated to the inorganic octahedral layers,[44, 45] it is interesting to investigate the influence of the organic moieties. Figure 2d depicts a similar temperature series of Raman spectra recorded from a single 2D layered perovskite flake with BA as the organic linker molecule. At room temperature, the Raman bands of $(BA)_2PbBr_4$ are less resolved with respect to the PEA system. Also, in this case, the Raman peaks become more defined with decreasing temperature, and at T=4K a series of sharp peaks can be resolved. Figure 3 shows the Raman spectra of $(PEA)_2PbBr_4$ and $(BA)_2PbBr_4$ flakes at low temperature (T=4K). The M1 band consists of a single peak for both BA and PEA systems, which is in line with its assignment to a twisting/rocking motion of the octahedra that should not depend significantly on the organic moiety.[39] Interestingly, the number of peaks for $(BA)_2PbBr_4$ is much smaller than for $(PEA)_2PbBr_4$, because the M2 and M3 bands split into fewer modes. The M2 band of the $(PEA)_2PbBr_4$ flake consists of a series of six closely spaced peaks, while that of the $(BA)_2PbBr_4$ flake has three separated peaks. The M3 band in the PEA flake is split into four peaks at 4K, while the M3 band for $(BA)_2PbBr_4$ remains a single peak that gains in intensity and sharpness with decreasing temperature. The M4 (gray shaded rectangles) and M5 (red-shaded rectangle at 125 - 135 cm$^{-1}$) bands that are broad at room temperature in both systems turn into two well-defined peaks at low temperature for both $(PEA)_2PbBr_4$ and $(BA)_2PbBr_4$ flakes. By comparing the positions of the vibrational bands (in frequency) that are indicated by the transparent coloured rectangles in Figure 3, we find that overall the vibrational frequencies are red-shifted in the BA system with respect to the PEA one. This behavior can be rationalized by the higher stiffness of the PEA molecules. Also, we find an overall blue-shift of all modes with decreasing temperature that is attributed to an increased rigidity of the bonds at lower temperatures. Since we do not observe any abrupt changes in the Raman spectra *versus* temperature, temperature-induced phase transitions can be excluded. Concerning systems with different halide anions, we would expect that their vibrational modes are qualitatively similar, and that the frequencies shift in relation to the size of the halide, in agreement with the lower frequencies reported for $(PEA)_2PbI_4$ in ref. [39].

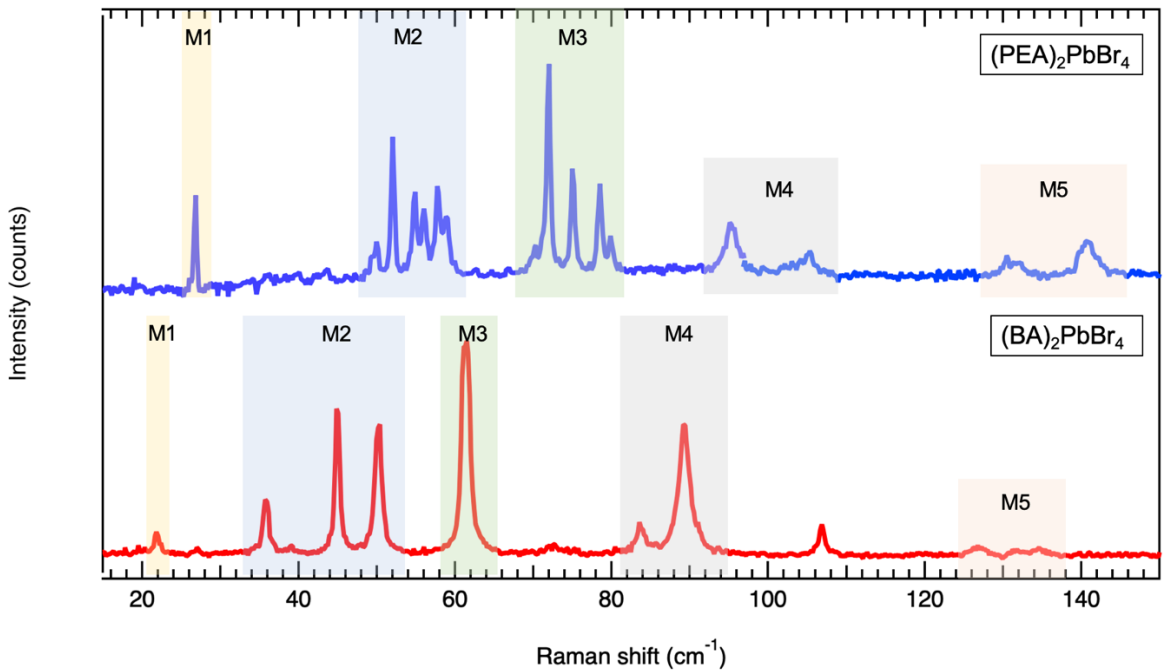

Figure 3. Ultralow-frequency Raman spectra of individual $(PEA)_2PbBr_4$ and $(BA)_2PbBr_4$ perovskite flakes recorded at T= 4K. The individual flakes are oriented with one main axis of the pseudocubic octahedral lattice parallel to the linear polarization of the excitation laser. Spectra are offset vertically for clarity. Peak positions are given in Table 1, and Figures S7 and S8 show Raman spectra recorded for an extended frequency range. The excitation wavelength was 633 nm.

To gain insight into the directionality of the vibrational modes, we analyzed the dependence of the Raman bands on the orientation of the flake with respect to the polarization of the excitation and collection signal. Since the exfoliated flakes are flat on the substrate surface, the linear polarization of the excitation and detection light has well-defined angles with respect to the orientation of the octahedral lattice, whose main pseudocubic axis orientations are evident by the straight edges of the flake. Figure 4 shows polarized and depolarized Raman spectra collected at T=4K for exfoliated flakes where one main axis of the pseudocubic octahedral lattice is either parallel (horiz.) or diagonal (45°) to the linear light polarization. The orientation of the flakes is shown in the insets, and the straight edges give the directions of the main axes of the pseudocubic octahedral lattice. For the horizontal flake direction, for both $(PEA)_2PbBr_4$ and $(BA)_2PbBr_4$ flakes almost all modes appear in the polarized spectra at 4K and are absent in the depolarized ones. This either indicates an isotropic character of the underlying vibrations, or implies that they are mainly in direction of the main pseudocubic axes. Only few weak peaks, for example at 33 cm$^{-1}$, 73 cm$^{-1}$ and 132 cm$^{-1}$ for the $(PEA)_2PbBr_4$ flake (Figure 4a) can be discerned in depolarized configuration, and these should have an intrinsic anisotropy. Interestingly, for a diagonal orientation (45°) of the flakes, the polarization-dependent Raman spectra of $(PEA)_2PbBr_4$ and $(BA)_2PbBr_4$ samples are very different. For $(PEA)_2PbBr_4$ flakes, the M2 and M3 bands show complementary behavior (Figure 4b), with the M2 band appearing only in the polarized spectrum, while the M3 band is present only in the depolarized one. This indicates a strong anisotropy that is induced by the PEA linker molecules, and the strong presence of the M3 band under 45° (see also Figure S9 for room temperature data) points to a configuration of the phenethyl rings along the diagonal directions of the pseudocubic octahedral lattice, as sketched in Figure 1b. On the contrary, for (BA)$_2$PbBr$_4$ flakes, where the inorganic layers are linked by linear alkyl chain molecules, all Raman peaks under the diagonal direction appear in both polarized and depolarized spectra (Figure 4d). Therefore, the BA linkers do not introduce any significant anisotropy.

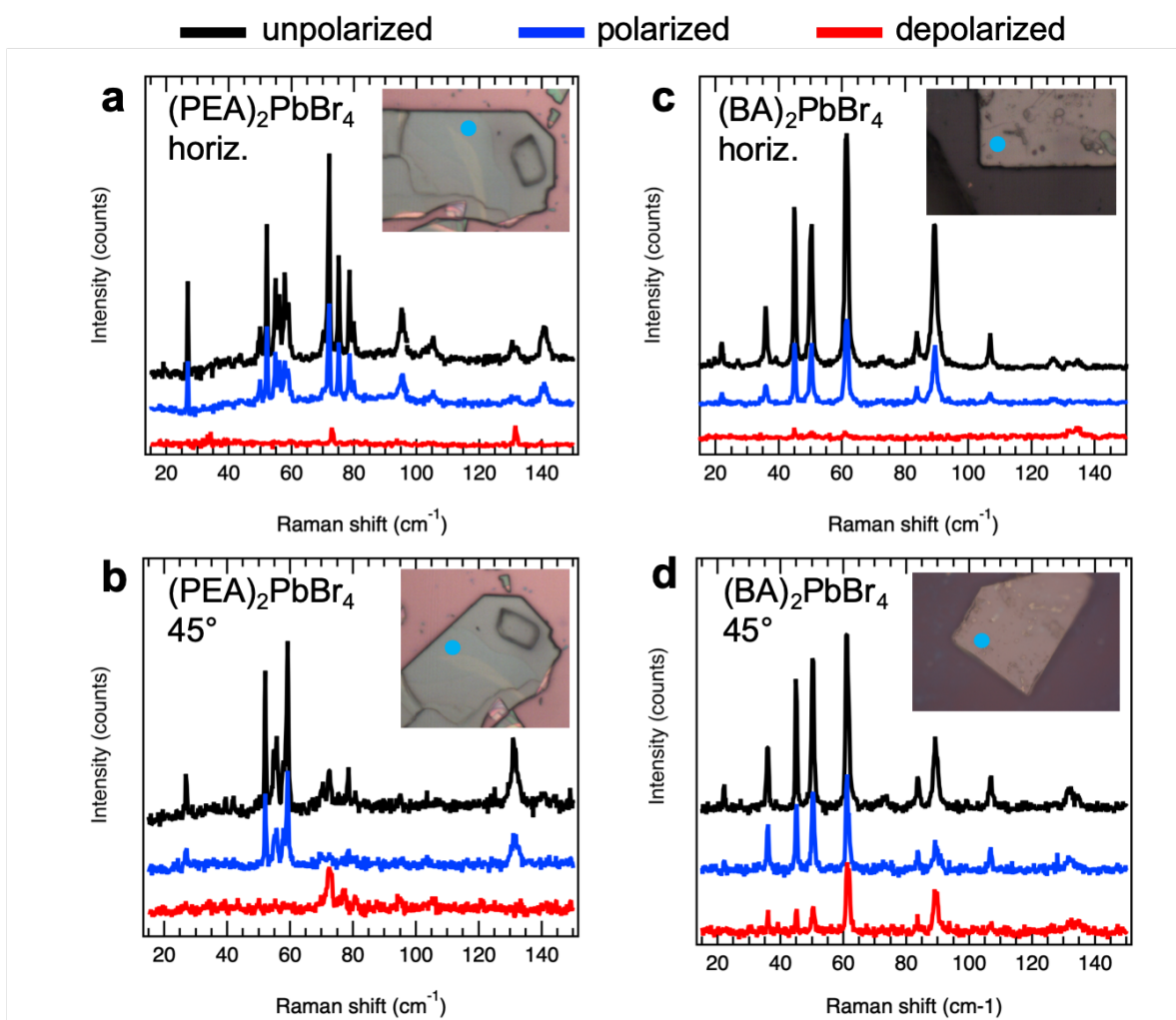

Figure 4. Polarization and orientation dependence of the vibrational bands. Polarized Raman spectra recorded from $(PEA)_2PbBr_4$ (a,b) and $(BA)_2PbBr_4$ (c,d) flakes at T=4K under laser excitation at 633 nm. Unpolarized spectra are plotted in black, polarized spectra with polarizer and analyzer in parallel direction in blue, and depolarized spectra with polarizer oriented perpendicular to the analyzer in red color. The insets show optical microscope images of the investigated flakes and their orientation, and the blue circles there indicate the positions chosen for collecting the spectra. The analyzer was kept in vertical direction with respect to these images.

The orientation dependence is also strongly present in the Raman spectra recorded at room temperature, as shown in Figure S9, where the M3 band shows strongest intensity in the depolarized spectra under 45° orientation, and is nearly absent in the polarized spectra under the same angle. For the horizontal direction, the M2 band is the dominant one, and M1 and M3 bands appear in the polarized spectra, but not in the depolarized ones. Interestingly the M5 mode has also a marked polarization dependence at room temperature, and is strongest in the depolarized spectrum under horizontal configuration.

Normal mode analysis was computationally performed for the $(PEA)_2PbBr_4$ and $(BA)_2PbBr_4$ structures in their crystalline form (taken from ref. [13]) after geometry relaxation at the gamma point. Periodic boundary conditions produce the *n=1* multilayer structures that correspond to the flakes that we investigated. To identify possible interlayer coupling effects we have also computed the modes of a single layer structure that was obtained by adding vacuum in the out of plane *(z)* direction. The calculated mode distributions are plotted in Figure S10, and display a very rich mode spectrum that makes it difficult to assign individual modes to the observed Raman peaks. However, the following can be concluded from the DFT modeling: In certain frequency bands the density of the computed modes is particularly high, for example around 40 cm$^{-1}$ and 70 cm$^{-1}$, and in these two bands the dominant vibrational motions are related to Pb-Br bond bending, and around 70 cm$^{-1}$ also some scissoring modes can be observed (see movie PEA_76cm-1.gif in the Supplementary Material) . For the (PEA)$_2$PbBr$_4$, these vibrations of the inorganic lattice go along with strong bending of the bulky and rigid PEA molecules and torsional motions of the phenethyl ring (see movies PEA_43cm-1.gif and PEA_59cm-1.gif in the Supplementary Material). Due to their smaller steric hindrance and their softer structure, the BA molecules in (BA)$_2$PbBr$_4$ multilayers perform all sorts of rotations and vibrations and therefore are not particularly in resonance with the vibrations of the inorganic octahedral lattice. Compared to the multilayer structures, the modes in the single layers are shifted to lower frequencies, which is reasonable, since one end of the molecules is free in this configuration. In particular, in the single layer (PEA)$_2$PbBr$_4$ structure, the bending of the N-C-C angle is released and the molecules unfold. Furthermore, the multilayer structures manifest a larger number of modes with respect to the single layers, which is more prominent for the (PEA)$_2$PbBr$_4$ system. The latter is in good agreement with our experimental Raman results and supports that mode splitting due to interlayer coupling occurs.

Based on the temperature, polarization and orientation dependent Raman spectra and our modeling results, we make a tentative assignment of the observed Raman peaks to the possible vibrational modes of the 2D layered perovskite flakes. The lowest frequency band, M1, can be associated to a rocking/twisting vibration of the heaviest subsystem, represented by the $[PbX_6]^{4-}$ octahedra, as already pointed out in other works on iodide based layered perovskites.[39] This octahedral vibration occurs along the main pseudocubic lattice directions, and should have one coupled phase across the octahedral plane, therefore it does not split. The next higher frequency vibrations should be dominated by Pb-Br bonds within the octahedra, and from modeling we find mostly Pb-Br bond bending oscillation in the frequency range from 30 to 80 cm$^{-1}$. It is therefore reasonable to relate the M2 and M3 bands to oscillations of these bonds. The M2 band is intense under excitation with linear polarization parallel to the main pseudocubic axes of the octahedral lattice in polarized configuration, while the M3 band is strong under an angle of 45° in the depolarized configuration. This behavior points to a vibrational motion along the two main pseudocubic axes for M2, and to vibrations along the diagonal of the lattice for M3. Here, the possible oscillations in the three different spatial directions of the 2D layered system could be the reason for the splitting of M2 into three modes, as it is observed for the (BA)$_2$PbBr$_4$ flakes. The additional mode splitting that occurs in the (PEA)$_2$PbBr$_4$ flakes, resulting in six peaks of the M2 band, is highly peculiar. It could originate from the ordered orientation of the phenethyl ring with respect to the octahedral lattice that induces directional anisotropy, or from vibrational coupling between adjacent octahedral layers that is enabled by the relatively rigid PEA linkers (but not by the softer BA ones). Concerning the M3 band, directionality along the diagonal of the octahedral lattice should play a strong role, and this could be induced by bending and scissoring of the in-plane Pb-Br, and Br-Pb-Br bonds, respectively. Also, in the case of the splitting of the M3 band into four peaks that is observed for (PEA)$_2$PbBr$_4$ flakes, the stiffness (and therefore interlayer coupling) and possibly an induced in-plane anisotropy of the PEA linkers could be the origin. The fact that M4 and M5 bands are much weaker and broader compared to M1-M3 points to damping induced by the organic moieties, which for Pb-Br bond stretching in the out-of-plane direction is reasonable. Here the splitting in two modes could result from in- and out-of-phase oscillations. The frequencies of the different peaks that were identified in the low-temperature Raman spectra, the symmetry of the modes from group theory analysis (performed on a single octahedron), and the assignment of the vibrational motion (from the discussion above) is summarized in Table 1. We note that even within this simple representation, we are able to assign an irreducible representation of the $D_{2h}$ symmetry group to the main vibrational motions. This assignment also agrees with those from ref. [44, 45]. However, the polarization dependent intensities that can be derived from the $A_g$, $B_{1g}$, $B_{2g}$ and $B_{3g}$ symmetry of the modes (see group theory discussion in the SI) in some parts does not agree with our experimental findings, which can be rationalized by the additional symmetry-breaking that is induced by the organic molecules.

Table 1. Frequencies of the vibrational modes discerned in the Raman spectra for (PEA)$_2$PbBr$_4$ and (BA)$_2$PbBr$_4$ flakes at T=4K, together with their symmetry, and the tentative assignment of the dominant vibrational motion.

| Band | (PEA)$_2$PbBr$_4$ freq. (cm$^{-1}$) | (BA)$_2$PbBr$_4$ freq. (cm$^{-1}$) | Irreducible representation ($D_{2h}$ symmetry) | Vibrational motion |
|---|---|---|---|---|
| M1 | 26.8 | 21.8 | $B_{1g}$, $B_{3g}$ | Octahedra rocking/twisting |
| M2 | 52.4 | 35.7 | $A_g$ | Pb-Br bond bending |
| | 54.9 | | | |
| | 56.0 | 44.9 | | |
| | 57.7 | 50.3 | | |
| | 58.8 | | | |
| M3 | 70.3 | 61.5 | $B_{2g}$ | Pb-Br bond bending and twisting; Br-Pb-Br scissoring in the octahedral plane |
| | 72.0 | | | |
| | 75.0 | | | |
| | 78.5 | | | |

|  | 80 |  |  |  |
|---|---|---|---|---|
| **M4** | 95.1 | 83.6 | $A_g$ | Out-of-plane Pb-Br bond stretching |
|  | 105.3 | 89.1 |  |  |
| **M5** | 131.6 | 106.8 | $A_g$ | In/Out-of-plane Pb-Br bond stretching |
|  | 140.7 | 132 |  |  |

Conclusions

We resolved the low-frequency vibrational modes in Ruddlesden-Popper 2D layered perovskite materials by polarized Raman spectroscopy on single flakes. Temperature and polarization dependent measurements allowed to group sets of modes in bands with common symmetry behavior. PEA and BA molecules as organic linkers led to different Raman bands, both quantitatively, as similar vibrational modes are shifted in frequency, and qualitatively, as modes with different symmetry are observed. In particular, PEA as organic moiety induced a distinct anisotropy in the vibrational bands that occur in the frequency range assigned to Pb-Br bond bending and scissoring (30 - 80 cm$^{-1}$), which should be related to the orientation of the phenethyl ring with respect to the inorganic octahedral lattice. Our detailed experimental characterization elucidates how organic moieties can be tailored for the design of the vibrational coupling between the organic and inorganic components in layered perovskite materials. Such knowledge is highly interesting for optomechanical and thermal coupling, where the interaction strength can be tuned by the choice of organic moiety and *via* the polarization direction of the incident light.

Methods

<u>Synthesis of layered Ruddlesden-Popper perovskites</u>. Two set of $RAm_2MA_{n-1}Pb_nBr_{3n+1}$, layered perovskite materials were prepared following our previously reported method.[14] Briefly, the selected amines (phenethylamine and butylamine) were added to a vial containing 95 mg of PbBr$_2$ dissolved in 60 µl of HBr and 1 mL of acetone. The resulting mixture was strongly stirred for few minutes and the perovskite crystals were collected and dried overnight on filter paper. We used

BA and PEA as ammonium ions ($RAm$), without a source of methylammonium (MA), which results in $(PEA)_2PbBr_4$ and $(BA)_2PbBr_4$ platelets with one octahedral layer ($n = 1$) in the stacks.

Exfoliation and transfer on Si-substrates. A deposit of platelets was placed on a non-emissive, one-sided 3M Scotch tape and the flakes were mechanically exfoliated by gently pushing a clean part of the tape onto the platelet deposits. The tape was then detached collecting thin flakes that were transferred to glass/Si substrates for the experiments.

Optical and morphological characterization. An MFP-3D AFM (Asylum Research) atomic force microscopy system operating in intermittent contact mode was used to image the exfoliated flakes and perform the topological analysis. The scan area was set as 5 μm × 5 μm for $(PEA)_2PbBr_4$ samples and 10 μm × 10 μm for $(BA)_2PbBr_4$ samples with a resolution of 256 × 256 pixels. A PANalytical Empyrean X-ray diffractometer, equipped with a 1.8 kW CuKα ceramic X-ray tube and a PIXcel3D 2x2 area detector, was used for X-Ray diffraction of the samples that were deposited on a zero-diffraction Si substrate. The diffractometer was operating at 45 kV and 40 mA. A Varian Cary 5000 ultraviolet–visible–near infrared (UV–vis–NIR) spectrophotometer equipped with an external diffuse reflectance accessory was used to collect the absorption spectra.

Raman spectroscopy. Raman experiments were performed in non-resonant conditions with wavelengths of 442 nm from a He-Cd laser and 632.8 nm from a He-Ne laser. 1800 lines/mm and 2400 lines/mm gratings were used in the Raman measurements, where the spectral resolution was 0.19 cm$^{-1}$ per CCD pixel under 632.8 nm excitation with 2400 lines/mm. The laser plasma lines were removed by Bragg-volume-grating-based bandpass filters from OptiGrate Corp. Measurements down to 5 cm$^{-1}$ for each excitation wavelength were achieved by three BragGrate notch filters from OptiGrate Corp with optical density of 3~4 with full width of half maximum of 5-10 cm$^{-1}$. Two vertical polarizers were used for the polarized and depolarized measurements, and a half wave plate was inserted to rotate the laser polarisation to be parallel (polarized) or perpendicular (depolarized) to the analyzer. The laser power was kept below 500 µW to avoid laser induced damage. In the temperature dependent measurements, the sample was first cooled to 4 K and then the temperature was raised stepwise to the indicated values.

The photoluminescence spectra were recorded with the Raman spectroscopy setup, using a grating with 600 lines/mm and excitation with a laser at 325 nm wavelength.

<u>Modeling.</u> DFT simulations were carried out using the CP2K software.[46] The experimental structures from ref. [13] were allowed a tight relaxation ($10^{-6}$ Bohr/Hartree threshold for forces, and $10^{-10}$ Hartree threshold for energy) to the ground state geometry. Goedecker-Teter-Hutter pseudopotentials and the MOLOPT basis set were employed within the Gaussian and plane waves (GPW) formalism. Group theory based on DFT calculations on a single $Cs_4PbBr_6$ octahedron in vacuum after geometry optimization using the ADF software were performed to analyze the symmetry of the Raman active modes. Cs was used as A-cation to keep the charge balance neutral and to reduce the computational costs.

**Supporting Information**. The following files are available free of charge.

The Supporting Information (PDF) contains:

Figure S1. X-ray diffraction spectra of the 2D layered perovskite materials.

Figures S2 and S3. Optical and atomic force microscopy images of exfoliated flakes.

Figure S4. Micro-PL spectra from single flakes.

Figure S5. Raman spectra of flakes with different thickness and extended frequency range.

Figures S6 – S8. Raman spectra of the perovskite flakes under ambient conditions and for extended frequency range, and the Raman signal of the substrate.

Figure S9. Polarized spectra recorded at room temperature.

Figure S10. Vibrational frequencies obtained by density-functional theory calculations.

Raman intensities derived from group theory analysis;

Videos showing the vibrational motion of the $(PEA)_2PbBr_4$ and $(BA)_2PbBr_4$ systems obtained by density functional theory calculations (GIF)

AUTHOR INFORMATION

**Corresponding Author**

*email: roman.krahne@iit.it.

**Author Contributions**

Yu-Chen Leng and Miao-Ling Lin performed the Raman experiments under guidance from Ping-Heng Tan and Roman Krahne. Balaji Dhanabalan and Milena Arciniegas fabricated the exfoliated perovskite flakes and performed and structural characterization, and Giulia Biffi and Ivan Infante calculated the vibrational modes by DFT. Roman Krahne, Milena Arciniegas and Liberato Manna coordinated the work. The manuscript was written through contributions of all authors. All authors have given approval to the final version of the manuscript. ‡These authors contributed equally.

ACKNOWLEDGMENT

The research leading to these results has received funding from the European Union under the Marie Skłodowska-Curie RISE project COMPASS No. 691185. Ping-Heng Tan acknowledges support from K. C. Wong Education Foundation and the National Natural Science Foundation of China (Grant Nos.11874350). BD and MA thank the Materials Characterization facility at the Istituto Italiano di Technologia for the technical support on XRD and AFM analysis.

REFERENCES

(1) Chen, Y.; Sun, Y.; Peng, J.; Tang, J.; Zheng, K.; Liang, Z., 2D Ruddlesden–Popper Perovskites for Optoelectronics. *Adv. Mater.* **2018,** *30*, 1703487.
(2) Thrithamarassery Gangadharan, D.; Ma, D., Searching for Stability at Lower Dimensions: Current Trends and Future Prospects of Layered Perovskite Solar Cells. *Energy Environ. Sci.* **2019,** *12*, 2860-2889.
(3) Wang, K.; Wang, S.; Xiao, S.; Song, Q., Recent Advances in Perovskite Micro- and Nanolasers. *Adv. Opt. Mater.* **2018,** *6*, 1800278.
(4) Mao, L.; Wu, Y.; Stoumpos, C. C.; Wasielewski, M. R.; Kanatzidis, M. G., White-Light Emission and Structural Distortion in New Corrugated Two-Dimensional Lead Bromide Perovskites. *J. Am. Chem. Soc.* **2017,** *139*, 5210-5215.
(5) Dou, L.; Wong, A. B.; Yu, Y.; Lai, M.; Kornienko, N.; Eaton, S. W.; Fu, A.; Bischak, C. G.; Ma, J.; Ding, T., Atomically Thin Two-Dimensional Organic-Inorganic Hybrid Perovskites. *Science* **2015,** *349*, 1518-1521.
(6) Cheng, B.; Li, T.-Y.; Maity, P.; Wei, P.-C.; Nordlund, D.; Ho, K.-T.; Lien, D.-H.; Lin, C.-H.; Liang, R.-Z.; Miao, X., Extremely Reduced Dielectric Confinement in Two-Dimensional Hybrid Perovskites with Large Polar Organics. *Commun. Phys.* **2018,** *1*, 80.
(7) Kamminga, M. E.; Fang, H.-H.; Filip, M. R.; Giustino, F.; Baas, J.; Blake, G. R.; Loi, M. A.; Palstra, T. T., Confinement Effects in Low-Dimensional Lead Iodide Perovskite Hybrids. *Chem. Mater.* **2016,** *28*, 4554-4562.
(8) Cao, D. H.; Stoumpos, C. C.; Farha, O. K.; Hupp, J. T.; Kanatzidis, M. G., 2D Homologous Perovskites as Light-Absorbing Materials for Solar Cell Applications. *J. Am. Chem. Soc.* **2015,** *137*, 7843-7850.
(9) Tsai, H.; Nie, W.; Blancon, J.-C.; Stoumpos, C. C.; Asadpour, R.; Harutyunyan, B.; Neukirch, A. J.; Verduzco, R.; Crochet, J. J.; Tretiak, S.; Pedesseau, L.; Even, J.; Alam, M. A.; Gupta, G.; Lou, J.; Ajayan, P. M.; Bedzyk, M. J.; Kanatzidis, M. G.; Mohite, A. D., High-Efficiency Two-Dimensional Ruddlesden–Popper Perovskite Solar Cells. *Nature* **2016,** *536*, 312.
(10) Pedesseau, L.; Sapori, D.; Traore, B.; Robles, R.; Fang, H.-H.; Loi, M. A.; Tsai, H.; Nie, W.; Blancon, J.-C.; Neukirch, A.; Tretiak, S.; Mohite, A. D.; Katan, C.; Even, J.; Kepenekian, M., Advances and Promises of Layered Halide Hybrid Perovskite Semiconductors. *ACS Nano* **2016,** *10*, 9776-9786.
(11) Mauck, C. M.; Tisdale, W. A., Excitons in 2D Organic–Inorganic Halide Perovskites. *Trends in Chem.* **2019,** *1*, 380-393.

(12) Leng, K.; Abdelwahab, I.; Verzhbitskiy, I.; Telychko, M.; Chu, L.; Fu, W.; Chi, X.; Guo, N.; Chen, Z.; Chen, Z.; Zhang, C.; Xu, Q.-H.; Lu, J.; Chhowalla, M.; Eda, G.; Loh, K. P., Molecularly Thin Two-Dimensional Hybrid Perovskites with Tunable Optoelectronic Properties Due to Reversible Surface Relaxation. *Nat. Mater.* **2018,** *17*, 908-914.

(13) Gong, X.; Voznyy, O.; Jain, A.; Liu, W.; Sabatini, R.; Piontkowski, Z.; Walters, G.; Bappi, G.; Nokhrin, S.; Bushuyev, O.; Yuan, M.; Riccardo, C.; McCamant, D.; Kelley, S. O.; Sargent, E. H., Electron–Phonon Interaction in Efficient Perovskite Blue Emitters. *Nat. Mater.* **2018,** *17*, 550–556.

(14) Dhanabalan, B.; Castelli, A.; Palei, M.; Spirito, D.; Manna, L.; Krahne, R.; Arciniegas, M., Simple Fabrication of Layered Halide Perovskite Platelets and Enhanced Photoluminescence from Mechanically Exfoliated Flakes. *Nanoscale* **2019,** *11*, 8334-8342.

(15) Stoumpos, C. C.; Cao, D. H.; Clark, D. J.; Young, J.; Rondinelli, J. M.; Jang, J. I.; Hupp, J. T.; Kanatzidis, M. G., Ruddlesden–Popper Hybrid Lead Iodide Perovskite 2d Homologous Semiconductors. *Chem. Mater.* **2016,** *28*, 2852-2867.

(16) Quintero-Bermudez, R.; Gold-Parker, A.; Proppe, A. H.; Munir, R.; Yang, Z.; Kelley, S. O.; Amassian, A.; Toney, M. F.; Sargent, E. H., Compositional and Orientational Control in Metal Halide Perovskites of Reduced Dimensionality. *Nat. Mater.* **2018,** *17*, 900-907.

(17) Straus, D. B.; Kagan, C. R., Electrons, Excitons, and Phonons in Two-Dimensional Hybrid Perovskites: Connecting Structural, Optical, and Electronic Properties. *J. Phys. Chem. Lett.* **2018,** *9*, 1434-1447.

(18) Yin, H.; Jin, L.; Qian, Y.; Li, X.; Wu, Y.; Bowen, M. S.; Kaan, D.; He, C.; Wozniak, D. I.; Xu, B., Excitonic and Confinement Effects of 2D Layered (C10h21nh3) 2pbbr4 Single Crystals. *ACS Appl. Energy Mater.* **2018,** *1*, 1476-1482.

(19) Long, H.; Peng, X.; Lu, J.; Lin, K.; Xie, L.-Q.; Zhang, B.-P.; Ying, L.; Wei, Z., Exciton-Phonon Interaction in Quasi-Two Dimensional Layered (Pea) 2 (Cspbbr 3) N-1 Pbbr 4 Perovskite. *Nanoscale* **2019**.

(20) Wang, K.; Wu, C.; Jiang, Y.; Yang, D.; Wang, K.; Priya, S., Distinct Conducting Layer Edge States in Two-Dimensional (2D) Halide Perovskite. *Sci. Adv.* **2019,** *5*, eaau3241.

(21) Zhai, Y.; Baniya, S.; Zhang, C.; Li, J.; Haney, P.; Sheng, C.-X.; Ehrenfreund, E.; Vardeny, Z. V., Giant Rashba Splitting in 2D Organic-Inorganic Halide Perovskites Measured by Transient Spectroscopies. *Sci. Adv.* **2017,** *3*, e1700704.

(22) Chen, Y.; Sun, Y.; Peng, J.; Zhang, W.; Su, X.; Zheng, K.; Pullerits, T.; Liang, Z., Tailoring Organic Cation of 2D Air-Stable Organometal Halide Perovskites for Highly Efficient Planar Solar Cells. *Adv. Energy Mater.* **2017,** *7*, 1700162.

(23) Spanopoulos, I.; Hadar, I.; Ke, W.; Tu, Q.; Chen, M.; Tsai, H.; He, Y.; Shekhawat, G.; Dravid, V. P.; Wasielewski, M. R., Uniaxial Expansion of the 2D Ruddlesden–Popper Perovskite Family for Improved Environmental Stability. *J. Am. Chem. Soc.* **2019,** *141*, 5518-5534.

(24) Lin, H.; Mao, J.; Qin, M.; Song, Z.; Yin, W.; Lu, X.; Choy, W. C., Single-Phase Alkylammonium Cesium Lead Iodide Quasi-2D Perovskites for Color-Tunable and Spectrum-Stable Red Leds. *Nanoscale* **2019,** *11*, 16907-16918.

(25) Smith, M. D.; Crace, E. J.; Jaffe, A.; Karunadasa, H. I., The Diversity of Layered Halide Perovskites. *Annu. Rev. Mater. Res.* **2018,** *48*, 111-136.

(26) Smith, M. D.; Karunadasa, H. I., White-Light Emission from Layered Halide Perovskites. *Acc. Chem. Res.* **2018,** *51*, 619-627.

(27) Zhou, G.; Li, M.; Zhao, J.; Molokeev, M. S.; Xia, Z., Single-Component White-Light Emission in 2d Hybrid Perovskites with Hybridized Halogen Atoms. *Adv. Opt. Mater.* **2019**, 1901335.

(28) Yang, Y.; Gao, F.; Gao, S.; Wei, S.-H., Origin of the Stability of Two-Dimensional Perovskites: A First-Principles Study. *J. Mater. Chem. A* **2018,** *6*, 14949-14955.

(29) Zheng, H.; Liu, G.; Zhu, L.; Ye, J.; Zhang, X.; Alsaedi, A.; Hayat, T.; Pan, X.; Dai, S., The Effect of Hydrophobicity of Ammonium Salts on Stability of Quasi-2D Perovskite Materials in Moist Condition. *Adv. Energy Mater.* **2018,** *8*, 1800051.

(30) Guo, P.; Stoumpos, C. C.; Mao, L.; Sadasivam, S.; Ketterson, J. B.; Darancet, P.; Kanatzidis, M. G.; Schaller, R. D., Cross-Plane Coherent Acoustic Phonons in Two-Dimensional Organic-Inorganic Hybrid Perovskites. *Nat. Commun.* **2018,** *9*, 2019.

(31) Arai, R.; Yoshizawa-Fujita, M.; Takeoka, Y.; Rikukawa, M., Orientation Control of Two-Dimensional Perovskites by Incorporating Carboxylic Acid Moieties. *ACS Omega* **2017,** *2*, 2333-2336.

(32) Castelli, A.; Biffi, G.; Ceseracciu, L.; Spirito, D.; Prato, M.; Altamura, D.; Giannini, C.; Artyukhin, S.; Krahne, R.; Manna, L., Revealing Photoluminescence Modulation from Layered Halide Perovskite Microcrystals Upon Cyclic Compression. *Adv. Mater.* **2018**, 1805608.

(33) Zhou, C.; Lin, H.; Tian, Y.; Yuan, Z.; Clark, R.; Chen, B.; van de Burgt, L. J.; Wang, J. C.; Zhou, Y.; Hanson, K., Luminescent Zero-Dimensional Organic Metal Halide Hybrids with near-Unity Quantum Efficiency. *Chem. Sci.* **2018,** *9*, 586-593.

(34) Chen, Y.; Yu, S.; Sun, Y.; Liang, Z., Phase Engineering in Quasi-2D Ruddlesden–Popper Perovskites. *J. Phys. Chem. Lett.* **2018,** *9*, 2627-2631.

(35) Yan, J.; Qiu, W.; Wu, G.; Heremans, P.; Chen, H., Recent Progress in 2D/Quasi-2D Layered Metal Halide Perovskites for Solar Cells. *J. Mater. Chem. A* **2018,** *6*, 11063-11077.

(36) Straus, D. B.; Hurtado Parra, S.; Iotov, N.; Gebhardt, J.; Rappe, A. M.; Subotnik, J. E.; Kikkawa, J. M.; Kagan, C. R., Direct Observation of Electron–Phonon Coupling and Slow Vibrational Relaxation in Organic–Inorganic Hybrid Perovskites. *J. Am. Chem. Soc.* **2016,** *138*, 13798-13801.

(37) Blancon, J.-C.; Stier, A. V.; Tsai, H.; Nie, W.; Stoumpos, C. C.; Traore, B.; Pedesseau, L.; Kepenekian, M.; Katsutani, F.; Noe, G., Scaling Law for Excitons in 2D Perovskite Quantum Wells. *Nat. Commun.* **2018,** *9*, 1-10.

(38) Ni, L.; Huynh, U.; Cheminal, A.; Thomas, T. H.; Shivanna, R.; Hinrichsen, T. F.; Ahmad, S.; Sadhanala, A.; Rao, A., Real-Time Observation of Exciton–Phonon Coupling Dynamics in Self-Assembled Hybrid Perovskite Quantum Wells. *ACS Nano* **2017,** *11*, 10834-10843.

(39) Thouin, F.; Valverde-Chávez, D. A.; Quarti, C.; Cortecchia, D.; Bargigia, I.; Beljonne, D.; Petrozza, A.; Silva, C.; Srimath Kandada, A. R., Phonon Coherences Reveal the Polaronic Character of Excitons in Two-Dimensional Lead Halide Perovskites. *Nat. Mater.* **2019,** *18*, 349-356.

(40) Yin, T.; Liu, B.; Yan, J.; Fang, Y.; Chen, M.; Chong, W. K.; Jiang, S.; Kuo, J.-L.; Fang, J.; Liang, P.; Wei, S.; Loh, K. P.; Sum, T. C.; White, T. J.; Shen, Z. X., Pressure-Engineered Structural and Optical Properties of Two-Dimensional (C4h9nh3)2pbi4 Perovskite Exfoliated Nm-Thin Flakes. *J. Am. Chem. Soc.* **2019,** *141*, 1235-1241.

(41) Taplin, F.; O'Donnell, D.; Kubic, T.; Leona, M.; Lombardi, J., Application of Raman Spectroscopy, Surface-Enhanced Raman Scattering (Sers), and Density Functional Theory for the Identification of Phenethylamines. *Appl. Spectrosc.* **2013,** *67*, 1150-1159.

(42) Teixeira-Dias, J. J. C.; de Carvalho, L. A. E. B.; da Costa, A. M. A.; Lampreia, I. M. S.; Barbosa, E. F. G., Conformational Studies by Raman Spectroscopy and Statistical Analysis of Gauche Interactions in N-Butylamine. *Spectrochim. Acta Part A* **1986,** *42*, 589-597.
(43) Polavarapu, L.; Nickel, B.; Feldmann, J.; Urban, A. S., Advances in Quantum-Confined Perovskite Nanocrystals for Optoelectronics. *Adv. Energy Mater.* **2017,** *7*, 1700267.
(44) Pérez-Osorio, M. A.; Lin, Q.; Phillips, R. T.; Milot, R. L.; Herz, L. M.; Johnston, M. B.; Giustino, F., Raman Spectrum of the Organic–Inorganic Halide Perovskite Ch3nh3pbi3 from First Principles and High-Resolution Low-Temperature Raman Measurements. *J. Phys. Chem. C* **2018,** *122*, 21703-21717.
(45) Pérez-Osorio, M. A.; Milot, R. L.; Filip, M. R.; Patel, J. B.; Herz, L. M.; Johnston, M. B.; Giustino, F., Vibrational Properties of the Organic–Inorganic Halide Perovskite Ch3nh3pbi3 from Theory and Experiment: Factor Group Analysis, First-Principles Calculations, and Low-Temperature Infrared Spectra. *J. Phys. Chem. C* **2015,** *119*, 25703-25718.
(46) Hutter, J.; Iannuzzi, M.; Schiffmann, F.; VandeVondele, J., Cp2k: Atomistic Simulations of Condensed Matter Systems. *WIREs Comput. Mol. Sci.* **2014,** *4*, 15-25.

**For Table of Contents Only**

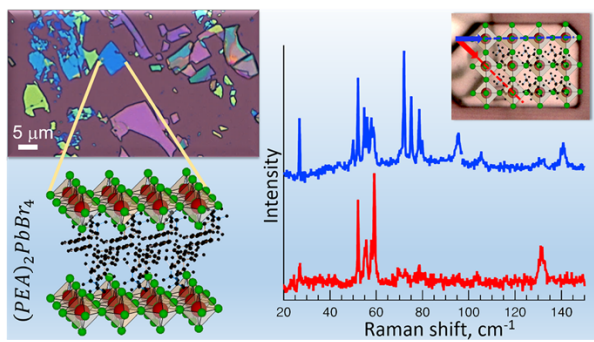